\documentclass[journal, doublecolumn, 10pt]{IEEEtran}
\usepackage{setspace}
\usepackage{amsmath,amssymb,bm,amsfonts,amsthm}
\usepackage{color}
\usepackage{algorithm,algorithmicx}
\usepackage[noend]{algpseudocode}
\usepackage{multirow}
\usepackage[dvips]{graphicx}
\usepackage{cite}

\newtheorem*{remark}{Remark}
\renewenvironment{IEEEbiography}[1]
  {\IEEEbiographynophoto{#1}}
  {\endIEEEbiographynophoto}

\makeatletter
 \let\old@ps@headings\ps@headings
 \let\old@ps@IEEEtitlepagestyle\ps@IEEEtitlepagestyle
 \def\confheader#1{%
 \def\ps@IEEEtitlepagestyle{%
 \old@ps@IEEEtitlepagestyle%
 \def\@oddhead{\strut\hfill#1\hfill\strut}%
 \def\@evenhead{\strut\hfill#1\hfill\strut}%
 }%
 \ps@headings%
 }
 \makeatother

\confheader{%
\parbox{20cm}{\small This work has been submitted to the IEEE for possible publications.\\
Copyright may be transferred without notice, after which this version may no longer be accessible.}
  }

\begin{document}

\title{SD-VEC: Software-Defined Vehicular Edge Computing with Ultra-Low Latency}
% Distributed Computing Architectures for Ultra-Low Latency Connected Transportation
\author{Shih-Chun~Lin, Kwang-Cheng Chen, and Ali Karimoddini %<-this % stops a space % , and Nicola Rohrseitz
\IEEEcompsocitemizethanks{
\IEEEcompsocthanksitem Shih-Chun Lin is with North Carolina State University; Kwang-Cheng Chen is with University of South Florida; Ali Karimoddini is with North Carolina A\&T State University.
%\IEEEcompsocthanksitem This work was supported by NC State 2019 Faculty Research and Professional Development (FRPD), Cisco Systems, Inc., North Carolina Department of Transportation (NCDOT), and Cyber Florida.
}% <-this % stops a space %\thanks{}
}
\markboth{ }%
\maketitle

\IEEEcompsoctitleabstractindextext{%
\begin{abstract} %% 7 pages $=>$ 5,500 words
New paradigm shifts and 6G technological revolution in vehicular services have emerged toward unmanned driving, automated transportation, and self-driving vehicles. As the technology for autonomous vehicles becomes mature, real challenges come from reliable, safe, real-time connected transportation operations to achieve ubiquitous and prompt information exchanges with massive connected and autonomous vehicles.
This article aims at introducing novel wireless distributed architectures that embed the edge computing capability inside software-defined vehicular networking infrastructure. Such edge networks consist of open-loop grant-free communications and computing-based control frameworks, which enable dynamic eco-routing with ultra-low latency and mobile data-driven orchestration. Thus, this work advances the frontiers of machine learning potentials and next-generation mobile system realization in vehicular networking applications.
\end{abstract}
}
\maketitle
\IEEEdisplaynotcompsoctitleabstractindextext
\IEEEpeerreviewmaketitle

\section{Introduction}\label{sec1}
Autonomous driving has attracted a considerable community interest, which drives the demand of intelligent vehicular systems and connected transportation to ensure ultra-low end-to-end latency. Such connected and autonomous vehicle (CAV) killer applications rely on fast mobile networking that keeps end-to-end communication latency down to 1 ms range for large-scale reliable deployment~\cite{USDOT2}. However, conventional communication protocol-based vehicular networking can only support the latency in the range of 20-100 ms for the following reasons. First, data transmissions rely on initial control signaling storms, which employ closed-loop physical-layer (PHY) protocols, like three-way handshaking, with thousands of control messages for power control or channel estimation in a single PHY link. Second, cumbersome and reactive protocol stacks and higher-layer optimization (i.e., routing and scheduling in networking) take a significant amount of time in computation and information gathering. Third, the high mobility of CAVs incurs tremendous overhead and complexity of protocols, which in turn causes considerable latency for reliable communications. Forth, cloud computation using remote data centers for heavy-weight computation can address the rapid growth of mobile applications and their diverse quality-of-service (QoS) requirements, but at the price of seconds or even larger order in end-to-end latency. Finally, cloud-based models require vehicular data to be transmitted to and stored in data centers, increasing chances of information leakage or loss and longer than expected latency which could even result in fatal accidents.

As one of 5G and beyond envisioned services, ultra-reliable and low-latency communications (URLLC) aim to provide secure data transmissions from one end to another with ultra-high reliability and deadline-based low latency~\cite{giordani2020toward}. Current 3GPP standards for URLLC require 1 ms hard latency over the air interface and 99.999\% system reliability, meeting the needs of autonomous vehicles for performing cooperation and safety functions. Since August 2018, normative works of new radio (NR) vehicle-to-everything (V2X) and enhanced URLLC have been launched in 3GPP Rel-16~\cite{3GPP_TR_38_886}. By December 2019, the 3GPP RAN Plenary meeting approved 24 new projects for 3GPP Rel-17 with one primary focus of bringing sidelink capabilities from automotive to smartphones and public safety.
Meanwhile, IEEE 802.11p, the basis for dedicated short-range radio communication (DSRC), establishes wireless access in vehicular environments (WAVE)~\cite{IEEE802.11p} and supports communications among vehicles to enable intelligent transportation services. This technology dedicated for connected vehicles helps exciting SAE Level 3 (conditional automation) trail and soon deployment and targets Level 4/5 (high/full automation) development for self-driving vehicles. However, technological efforts to support mission-critical URLLC requirements are still in an early research stage with no concrete solutions yet.

\subsection{Edge Computing and Software-Defined Paradigms}\label{sec2}
Edge computing, a new computing paradigm, provides an alternative to eliminate the reliance on remote data centers for mobile applications. By moving the computing resources to the proximity of data sources, edge computing enables large amount of real-time and low-latency big data processing~\cite{ndikumana2019joint}. This paradigm removes the network bandwidth constraint to send all the data back and forth to the cloud. Despite of its innovative ideas, edge computing in vehicular networking is a challenging and daunting task. First, the underlying vehicular infrastructure consists of highly mobile and unreliable wireless links, hindering the network connectivity.
Second, efficient deploying edge computing nodes is difficult as system architects can over-estimate the capability of distributed processing and cause the unexpected failures.
Finally, the overhead of scaling applications into distributed edge nodes, coordinating cross-node state and storage, and handling inconsistent state delicately can also annihilate the benefit of edge computing.

Recently, software-defined networking (SDN) has been introduced to facilitate system design and management flexibility, which can potentially solve the challenges in vehicular edge computing. The main ideas of SDN are (i) to separate the data forwarding plane from the network control plane and (ii) to introduce novel network control functionalities based on a network abstraction. Hence, SDN provides the upper-running applications with a centralized view of distributed network states. This in turn can dramatically improve network resource utilization, simplify network management, reduce operating cost, and promote innovation and evolution. An overview of SDN-enabled Internet of vehicles (IoVs) is presented in~\cite{zhuang2019sdn}, which leverages SDN technologies in communication, computing, and caching to enhance vehicular system performance.
However, current SDN realizations mainly focus on programmable packet forwarding via flow tables and wildcard rules. The technology innovation and practical deployment of distributed computation and learning techniques with underlying vehicular infrastructure remain limited in literature.

\subsection{SD-VEC: Software-Defined Vehicular Edge Computing}
The objective of this article is to introduce novel wireless distributed architectures, namely software-defined vehicular edge computing (SD-VEC), that employs edge computation to enable ultra-low latency vehicular services, particularly for connected transportation.
SD-VEC seamlessly integrates ``intelligence" (i.e., computation capability) into anchor nodes (ANs), which connect roadside access points (APs) via backbone networks (e.g., cellular core networks and Internet).
The ANs can explore several computing operations at the vehicular network edge while establishing minimum one-time signaling via open-loop transmissions and proactive decisions between vehicles and the infrastructure (i.e., APs).
SD-VEC extends the network abstraction and provides an interface to allow intelligent ANs to control data delivery and processing.

To fill the gap for ultra-low latency mobile applications, SD-VEC replaces protocol-based vehicular networking with computing-based architectures that support edge computation near data sources.
The main contributions of SD-VEC lie in the following aspects.
First, SD-VEC offsets communication workloads to computation tasks, reducing networking latency while preserving reliability.
Second, ANs can respond to vehicles earlier via distributed processing without going through the remote cloud, mitigating total traffic going through the network.
Third, with edge computation and machine learning techniques, high vehicle mobility and subsequent impacts can be completely resolved by proactive access/association and anticipatory management~\cite{lin2018anticipatory}.
Finally, ANs can pre-process or filter sensitive vehicular data, making SD-VEC less fragile to security issues with enhanced data privacy.
In addition, as SD-VEC offloads computation to the vehicular infrastructure, it facilitates on-board computing in autonomous vehicles.

The rest of the paper is organized as follows. Section~\ref{sec3} introduces the software-defined computing infrastructure of SD-VEC. Section~\ref{sec4} and Section~\ref{sec5} provide ultra-low latency connected networking and big data-enabled orchestration for SD-VEC, respectively. Finally, Section~\ref{sec6} concludes the paper.

\section{Software-Defined Vehicular Computing Infrastructure}\label{sec3} %% 1.5 pages
\begin{figure}[!t]\centering
\includegraphics[width=3.2 in]{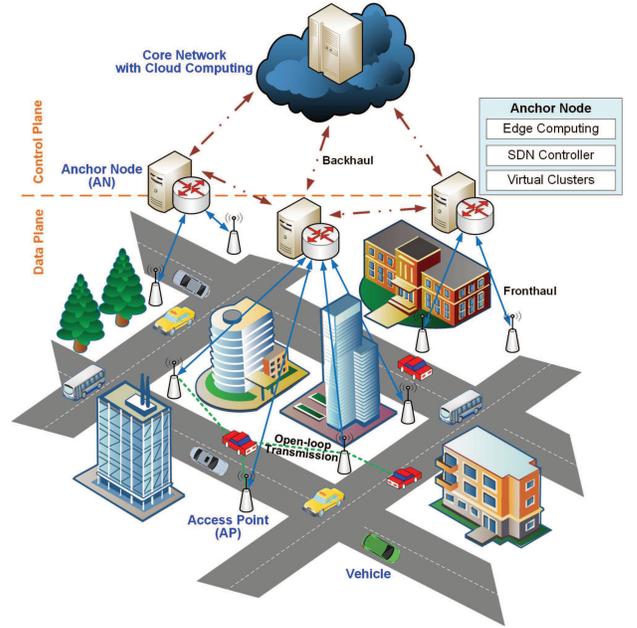}
\caption{The system infrastructure of SD-VEC.}\label{fig:Cloud-RAN1}
\end{figure}

This section describes the SD-VEC infrastructure, as shown in Figure~\ref{fig:Cloud-RAN1}, which allows vehicular applications to leverage the concepts of edge computing and SDN for ultra-low latency networking.
The vehicular data plane is an open, programmable, and virtualizable network forwarding infrastructure for endpoint traffic through last-mile open-loop transmissions. It consists of hardware-only APs and software-implemented virtual clusters/networks in ANs, which are connected via fronthaul links using standardized interfaces, such as common public radio interface (CPRI).
As ANs (or virtual clusters) govern the functionalities of APs, SD-VEC employs this distributed transmit point structure, which offers excellent cooperative gain and evolvability by aggregating massive technology-evolving APs at the ANs.

The vehicular control plane, handled by ANs and connecting to cloud servers via backhaul, has two components: (i) network management tools, such as SDN controllers and virtual cluster orchestration for traffic monitoring and real-time control/management of vehicular operation, and (ii) customized applications of service providers, like security and privacy solutions. These components can be designed, deployed, and updated to fit the specific and ever-changing needs. By equipping these ANs with the edge computing capability, a suite of innovative computation tools can be realized in SD-VEC to bolster ultra-low latency networking, including automatic task allocations between edge ANs and the cloud, anticipatory mobility management, and data analytic assisting predictive safety and real-time actions.
Our preliminary results on online automatic tasking can be adopted for safe coordination of actions taking into account the operation cost and time of the actions and scenarios for commutation network management and traffic control~\cite{AK2}. % AK1,

\subsection{Wireless Grant-Free Data Plane}\label{wgrantfree}
Serving as the foundation of SD-VEC infrastructure, an ultra-fast vehicular data plane with high reliability is developed, which exploits latency-minimum open-loop transmissions, extends multiuser detection techniques to support contention-free wireless access, provides multi-path based error control mechanisms to ensure reliability, and supports the potential of dynamic radio resources assignments among virtual clusters.
To facilitate proactive access between vehicles and APs, the ANs in SD-VEC employ the edge computing with mobility management and resource allocation.
Extended from our prior study~\cite{TWCdownlink2020}, a single vehicle can be treated as the center node of a virtual AP cluster so that multiple APs serve this vehicle via cooperative communications. This cell-free/virtual-cell concept is different from conventional protocol-based architectures where a cell/cluster uses one base station to serve multiple vehicles. We list the operations of proactive open-loop resource allocation as follows.
\begin{itemize}
\item Non-orthogonal multiple access (NOMA) and real-time beamforming: grant-free NOMA is adopted for PHY transmissions~\cite{shahab2020grant}. A contention transmission unit (CTU) as the basic multiple access resource consists of radio resources (e.g., time-frequency blocks), reference signal (e.g., demodulation reference signal, preamble), spreading sequence (e.g., codebook/codeword, sequence), etc.. Vehicles perform such contention-based access by randomly selecting CTUs or following pre-configured settings. To enable fast access in multiple antenna scenarios, unsupervised learning methods can be also adopted to provide
    beamforming operations in real-time (see, e.g.,~\cite{chbeamforming}) by avoiding compute-intensive labeling effort in training.

\item Proactive resource allocation: vehicles proactively allocate radio resources for their uplink transmissions in a distributed manner. For example, each vehicle can select its serving APs by piggybacking signal qualities and randomly allocating physical CTUs to each associated AP. A more sophisticated design is to apply iterative learning between uplink and downlink access. Initial uplink resource allocation can be determined via random decisions. Based on uplink results, downlink access is achieved through machine learning techniques in Section~\ref{sec3b}. Next-round uplink access can further learn from previous downlink allocations.

\item Multi-path cooperative communication: each vehicle packs its data into the selected CTU and transmits it to proactively associated APs. The size of associated AP implies the number of cooperative paths in the uplink. Each AP behaves as a relay node in cooperative multi-path networking (see, e.g.,~\cite{lin2015cognitive}) to forward the received signals to an AN.
    Amplify-and-forward can be selected as a cooperative communication strategy, which allows the AN to decode the vehicular data by using selection-combining with proactively selected CTUs. Decode-and-forward can be an alternative, by which each AP demodulates and decodes before relaying signals to the AN.

\item Error control mechanism: since there is no feedback with proactive open-loop communications to shorten the networking latency, multiple networking paths will be employed to ensure reliability and physical error-rate performance. Particularly, path-time codes will be developed as the error control coding over open-loop multi-path transmissions. Moreover, forward error correcting codes can still be applied to open-loop PHY communications. According to our preliminary study, low density parity check codes (LDPC) well serve this purpose with small-length packets, i.e., 128 or 256 bits packet payload.

\end{itemize}
As proactive resource allocation proceeds in a distributed manner, multi-armed bandit methodology can be adopted to strike the tradeoff between reliability and throughput.

\subsection{Machine Learning-Enhanced Network Architecture}\label{sec3b}
\begin{figure}[!t]\centering
\includegraphics[width=3.2 in]{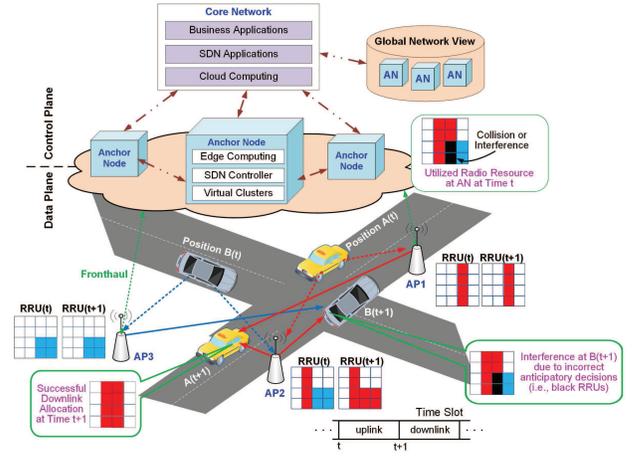}
\caption{Proactive uplink and downlink access.}\label{MACC1}
\end{figure}

\begin{figure}[!t]\centering
\includegraphics[width=3.6 in]{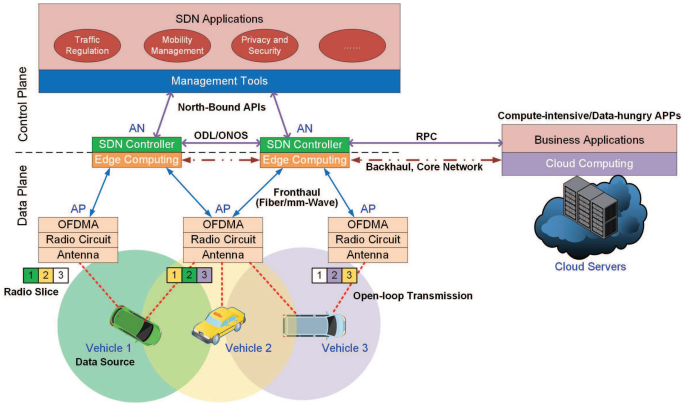}
\caption{Distributed computing architecture with a scalable control plane.}\label{fig:newCloud-RAN}
\end{figure}

Downlink resource allocation to support multiple vehicles via different APs can be realized in a centralized manner. However, different from conventional centralized management and operation, open-loop proactive downlink access relies on uplink access and precise anticipatory mobility management.
In particular, due to distributed decisions by vehicles, proactive uplink access inherits probable collisions and cannot provide useful transmissions without further sophisticated designs.
Figure~\ref{MACC1} provides an example of proactive uplink and downlink access in vehicular networking, where amplify-and-forward scheme is adopted and multiple access interference (MAI) happens in the overlapped CTUs (i.e., black blocks) for downlink access.
To address this issue, we introduce a software-defined computing architecture upon SD-VEC with  distributed machine learning-enhanced operations, as in Figure~\ref{fig:newCloud-RAN}. This architecture refers to URLLC of 5G and beyond and bandwidth reservation and downlink migration
from IEEE 802.11ax in the air-interface, which extends DSRC for next-generation vehicular services~\cite{TWCdownlink2020}. By exploiting agent and edge computation capability in the architecture, necessary operations for last-mile transmissions are designed to yield desired system performance while combating time-varying MAI, as listed below:
\begin{itemize}
\item Multiuser detection for downlink receivers: an effective technique to tackle the downlink MAI is developed by applying maximum-likelihood multiuser detection techniques, which can demodulate collided dowlink transmissions and provide each downlink signal separately. While the techniques might incur a high computational complexity, computation-efficient schemes can be developed for vehicles given that only few APs will be possibly involved due to moderate uplink ranges.

\item Anticipatory mobility management \& intelligent resource allocation: although correct downlink prediction cannot be always guaranteed such as in Figure~\ref{MACC1}, precise anticipatory management is critical for open-loop proactive communications, which determines potential APs in the next downlink time instant to serve vehicles.
    We introduce a powerful data-driven scheme by using historical data as shown in Section~\ref{ddamm}. Besides, through SDN construction and its capability of centralized control, an intelligent resource allocation can be further applied to optimally assign downlink associations (i.e., network slices) and the corresponding CTUs (i.e., radio slices) with details in Section~\ref{sec4}. This smart allocation significantly improves the throughput and reliability in proactive downlink.

\end{itemize}
Based on the designed architecture, %with distributed machine learning capability,
SDN controllers in ANs can execute desirable online management policies %(e.g., network association, resource allocation, and corresponding network protocols)
to ensure maximum system capacity for proactive multiple access.

\subsection{Scalable Software-Defined Control Plane}
This section describe the essential network management tools of SD-VEC that provide the capability of real-time control of vehicular operations with evolvability and adaptiveness. Particularly, an effective, scalable control plane is designed to ensure unified network status and control/computing coordination among distributed ANs.
\begin{itemize}
\item Optimal placement of multiple controllers: %vehicle location is time-varying, and the controller placement in SD-VEC should also consider distributed processing frameworks from edge computing.
    the optimal number of placement problems in SD-VEC aims to minimize the required SDN controllers (to reduce infrastructure cost for efficient placement) and simultaneously support all control traffic from vehicles (to ensure a feasible solution for scalable placement). Through the models of vehicle mobility and computing capability of ANs, the multi-controller placement determines (i) the number of required controllers, (ii) their geo-locations, and (iii) the control domain assignments for each vehicle.

\item Mobility-aware control traffic regulation: to promise on-line and adaptive traffic engineering in SD-VEC, we develop a novel dynamic control traffic balancing that decides the optimal forwarding paths of control flows with respect to mobile traffic. Dynamic traffic behaviors are modeled, and a nonlinear optimization problem for control traffic balancing is formulated to find the optimal forwarding paths for vehicles while minimizing the average control traffic latency. The obtained latency values are further feedbacked to the placement decision and trigger the adaptive control for better controller placement and timely control message delivery.

\item Distributed controller consensus: inter-controller synchronization should synchronize both controllers' network and computing-wide views to keep the network control logic centralized and to localize all decisions to each controller for minimizing control plane response time. We develop a distributed multi-controller synchronization that converges timely by exploiting the controller's network topology via the graph theory.

\end{itemize}

\section{Ultra-Low Latency Connected Vehicular Networking}\label{sec4} %% 1.5 pages
Emerging and growing vehicular applications require highly differentiated networking and computing capabilities to be integrated and deployed over the same vehicular infrastructure. To supports infrastructure-as-a-service for applications and latency-minimum association between vehicles and the infrastructure, we develop virtual cluster orchestration and vehicular network virtualization. In each virtual cluster/network, a vehicular application is provided with the ability to control, optimize, and customize the underlying resources without interfering other services' performance, leading to cost-efficient operations and enhanced QoS. We describe our designs of dynamic clustering and edge network slicing in the following.

\subsection{Latency-Optimal Vehicle-Centric Clustering}
\begin{figure}[!t]\centering
\includegraphics[width=3.4 in]{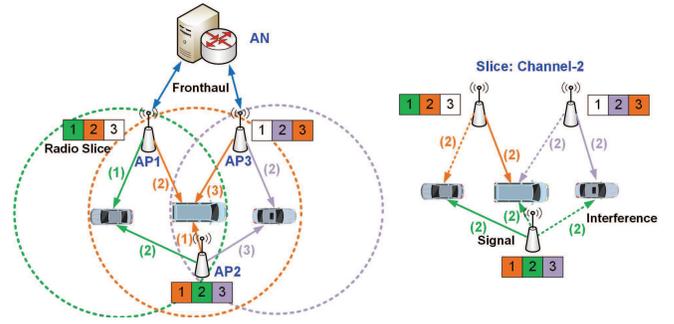}
\caption{Virtual clusters with radio resource slices.}
\label{vcc}
\end{figure}
As shown in Figure~\ref{vcc}, by forming vehicle-centric virtual clusters within SD-VEC, each vehicle is served by a selected subset of neighboring APs via open-loop transmissions and coordinated multipoint techniques.
Such a distributed transmit point architecture via fronthaul links between an AN and a cluster of APs can significantly boost the coordination gain of APs, eliminate cluster edges from severe inter-cluster interference, and in turn reduces vehicular networking latency. While the vehicle-centric clustering is challenging since the clusters are chosen in a dynamic way and may overlap, such a clustering is preferred when dealing with mobile objects.

We implement this virtual cluster orchestration via distributed reinforcement learning techniques for vehicular edge networks~\cite{vehicularedge2}. The designed orchestration jointly optimizes power allocation and vehicle association in the highway transportation, where each vehicle associates with a dedicated AP cluster for latency-minimum networking. Results show that our design can yield near-optimal performance with high reliability by a nominal number of training episodes than the baseline. This preliminary work establishes the feasibility of learning-enabled cluster optimization via edge computing capability to obtain remarkable efficiencies and rapid network operation.

\subsection{Eco-Vehicular Edge Network Slicing}
To achieve wireless virtualization in highly mobile environments, a comprehensive set of resources (e.g., antenna elements, APs, APs to the AN connections, wireless spectrum, transmission power, and ANs' computing capabilities) and the correspondingly slicing mechanism should be investigated with regard to time-varying factors (e.g., vehicle mobility, channel fading). Network slicing frameworks partition multi-dimensional wireless and computing resources into non-conflicting virtual networks, while the requests of each virtual network can be satisfied with the minimum operational expense. Accordingly, we have developed eco-vehicular edge networks for connected transportation~\cite{vehicularedge1}, which uses distributed multi-agent reinforcement learning to combat power-hungry edges while assuring system reliability and data rates. The designed ``dynamic eco-routing" makes ANs with respective edge Q-learners can collaboratively yield best-reward decisions for optimal downlinks. Evaluating by 3GPP cellular-based V2X services~\cite{3GPP_TR_38_886}, our solution outperforms conventional schemes in energy efficiency and system reliability and coverage. The virtual network framework also serves the security of virtual networks in Section~\ref{vesecurity}.

\section{Big Vehicular Data-Enabled System Orchestration}\label{sec5} %% 1.5 pages
Open-loop communications and virtual AP clustering form the core technologies in SD-VEC to accomplish ultra-low latency vehicular networking with high reliability. In this section, we extend intelligent edge computing at ANs by leveraging big vehicular data and distributed data processing. Big vehicular data provides unprecedented opportunities for system architects (or SDN controllers) to understand the requirements and behaviors of vehicle mobility and various network elements. With both instantaneous and historic data, big data-driven optimization allows intelligent real-time decision making in a wide range of services.
Useful features (e.g., the correlation between vehicular events and data traffic) can be extracted to make optimal decisions on (i) data/resource allocation and caching, (ii) anticipatory mobility management, and (iii) system security and privacy issues based on long-term strategies. And accordingly, deployment and operational costs are significantly minimized with improved network efficiency.

\subsection{Joint Data Caching and Computation Offloading}
\begin{figure}[!t]\centering
\includegraphics[width=3.2 in]{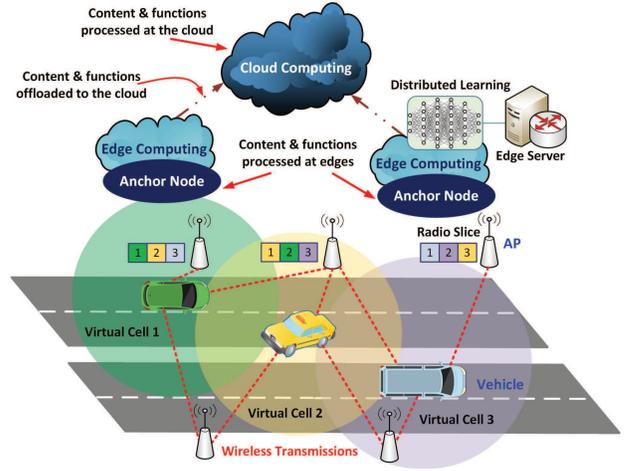}
\caption{ANs can cache and offload data and services between vehicular edge computing and cloud computing.}\label{fig:offloading}
\end{figure}
Edge computing at ANs can push the frontier of computing applications, data, and services away from centralized cloud computing infrastructures to the logical edge of vehicular networks, thereby enabling analytic and knowledge generation to occur closer to data sources. As shown in Figure~\ref{fig:offloading}, ANs, endowed with edge servers for cloud-like computing and storage capacity, can serve vehicular data as a substitute for the cloud. Extra tasks exceeding the ANs' computing capacity are further offloaded to the cloud, resulting in a hierarchical offloading structure between ANs and the cloud.

We have established application-defined networks that enable edge resource orchestration and adaptivity in terms of desired QoS for mobile AI applications through the collaboration with Cisco Systems. We extend with vehicular edge networks and investigate two typical edge-cloud operations: caching and offloading for contents and services. First, caching function refers to caching services and related databases/libraries into edge ANs, enabling vehicular data to be processed locally. Due to limited computing and storage resources at ANs, services cached on ANs determine tasks to be offloaded to the cloud, thereby significantly affecting the edge computing performance. Second, computation offloading operation concerns what/when/how to offload vehicular workload from local ANs to the cloud regarding the service availability at ANs (i.e., what types of computation tasks/applications). The optimal offloading decisions are complex and coupled both spatially and temporally because of high mobility in vehicular networks. Hence, an efficient solution is needed to optimize service caching and task offloading jointly for vehicular data and applications. Upon open-loop PHY and virtual clusters, the objective is to minimize the computation latency under a long-term energy consumption constraint.

\subsection{Data-Driven Anticipatory Mobility Management}\label{ddamm}
Anticipatory mobility management aims at predicting the proactive network association (i.e., Section~\ref{wgrantfree}) in the future by using machine learning techniques.
Specifically, instead of using GPS information of vehicles (which consumes huge communication bandwidth and jeopardizes privacy and security), data analytic at ANs can use the past history of AP association vectors to predict the AP association at next time slot. This data-driven framework simply avoids probing complicated, dynamic channel states, and losing prediction accuracy due to inaccurate GPS data.

Our recent study in~\cite{lin2018anticipatory} introduces such a data-driven mobility management scheme via three designs including (i) online association learning, (ii) mobility pattern assisted mechanism, and (iii) transferring learning for reinforcing mobility management. First, with a Markovian system model, the posterior belief of vehicle locations can be inferred from historical association vectors and vehicle velocity estimation. Accordingly, the association vectors can then be obtained via recursive Bayesian estimation.
Second, fixed road structure limits the moving patterns of vehicles with shared similarity. Data-driven frameworks can exploit this mobility pattern to improve the predication accuracy. Particularly, by modeling vehicle mobility with a Markov jump process, we can acquire a more accurate prediction of current vehicle locations based on the past trajectory, which is conditioned on the corresponding series of road geographic maps.
Finally, in addition to applying empirical probability for predicting mobility patterns, deep learning techniques with big vehicular data can facilitate more robust mobility management.
For example, random forest technique can be applied to reinforce anticipatory mobility management. The procedures of such edge analytic are (i) selecting data information (i.e., decision trees), (ii) potentially combining data decision trees, and (iii) extracting useful features via deep learning.
%This offline learning with historical data will greatly benefit emerging ITU-T ML5G standards.

\subsection{Virtualization-Enabled Security and Data Privacy}\label{vesecurity}
\begin{figure}[!t]\centering
\includegraphics[width=3.4 in]{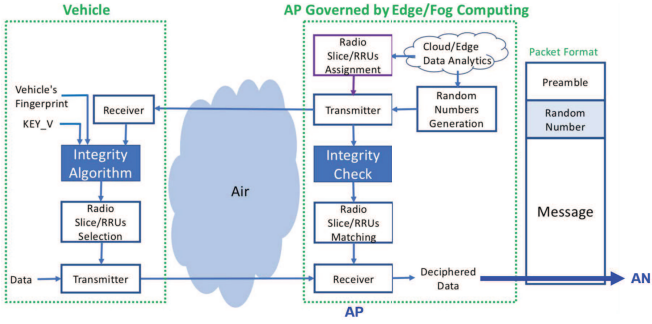}
\caption{Fingerprint-based dynamic stream cipher between a vehicle and an AN.}\label{security}
\end{figure}

Based on virtual clusters and resource slicing in Section~\ref{sec4}, SD-VEC can realize vehicle infrastructure-as-a-service to greatly facilitate system security and data privacy.
In particular, SD-VEC can decouple network security functions (e.g., fire-walling) from proprietary hardware appliances to run services in software and deliver network security enforcement by separating the security control plane from the security processing and forwarding planes. We propose a light-weight dynamic cipher that strengthens the security of symmetric stream cipher (i.e., a sequential uplink/downlink data transmissions) between vehicles and edge infrastructure, while retaining the desired ultra-low latency networking. First, a fingerprint-based dynamic stream cipher in Figure~\ref{security} is developed, where a vehicle's fingerprint is a nonlinear time-dynamic function and can be shared only between the vehicle and edge ANs. A simple fingerprint is vehicle's historical connections to APs, which is unique to each vehicle and aware by ANs.
Next, this fingerprint-based cipher can further exploit proactive resource allocation and edge analytic to secure the entire SD-VEC. Detailed procedures are given as follows.
\begin{itemize}
\item Authentication: dynamic stream cipher can seamlessly integrate with latest authentication schemes in mobile communications without requiring real-time authentication. After the authentication, ANs will send a starting key to start the cipher.

\item Fingerprint generation and verification: while the association attempt to APs might not be successful due to a failure/inaccuracy caused by MAI, ANs can still be aware of possible connected APs for each vehicle at any given time through the scalable SDN control plane. Hence, an integral checking algorithm is developed to verify the key and also use for downlink transmissions. ANs will maintain a master key for any loss of distributedly generated key by vehicles.

\item Edge network dynamic and fault tolerance: this cipher scheme can exploit the features of proactive network association, dynamic resource slicing, and cooperative communications with APs and provide another dynamic randomness to attackers. Resilient fault diagnosis and accommodation can also be employed for both network management and vehicle control which is robust to failures in network, edge, or vehicles operation~\cite{AK3}.

\end{itemize}

For the data privacy, we develop federated multi-task learning techniques over ANs with decentralized vehicular data. Since all the computation and data analytic happen at edge, ANs can filter or pre-process sensitive data information while releasing the computation burden of vehicles. This approach avoids information leakage in backhaul and cloud servers and thus makes SD-VEC less fragile to privacy issues. Besides, different from conventional mobility management, our anticipatory mobility management at ANs does not rely on private GPS data of vehicles but employ edge computing to predict plausible APs for proactive access by vehicles. Edge ANs only need to know the past vehicle traces (i.e. virtual clustering patterns that show the past association between vehicles and APs), which completely eliminates the privacy concern of using private location information.

\begin{remark}[NC-CAV Center]
Funded by the North Carolina Department of Transportation (NCDOT) in February 2020, the NC Transportation Center of Excellence on Connected and Autonomous Vehicle Technology (NC-CAV) leads cutting-edge research and development of CAV technologies and their impacts on transportation systems in NC and the nation. As core investigators in the NC-CAV, we are implementing this paper's solutions for smart and connected infrastructures in real-world CAV deployments.
By collaborating with Verizon Wireless for its 5G access, we are conducting an experimental validation to assess NC's readiness for CAVs in traditional and emerging infrastructure needs. The entire field deployment and testing in Greensboro will be accomplished by the Q2 of 2022.
\end{remark}

We are currently building an in-lab experimental testbed for SD-VEC architectures in Figure~\ref{fig:newCloud-RAN}. It is a fully-functional prototype for an SDN-based edge network and computing platform that consists of Odroid-C2 and Raspberry Pi 3 Model B as seven mobile data sources with computation workloads, TL-WDR4300 as five programmable APs, Gbit Ethernet as fronthaul links between APs and ANs, and Intel NUC7i5BNK as high-performance integrated storage servers at two ANs and a central cloud. We are validating our open-loop communication designs, grant-free proactive access, edge computation capability with Ryu controllers, and optimal real-time reconfiguration of mobile networking and computing functionalities. The results show that with 128 MB data chuck and Hadoop application, the small-scale testbed can provide a 53\% reduction in average read and write time. We will further examine this platform with latency-sensitive interactive gaming and expand it to enable larger-scale research.

\section{Conclusion}\label{sec6}
Ultra-low latency connected transportation is envisioned as a critical 6G vertical that will reshape autonomous driving and human society in the next five to 10 years.
Major development efforts in vehicular edge computing and wireless network infrastructures are facilitating the practical deployment of massive CAV technologies.
In this paper, we have presented distributed computing architectures and highlighted innovative solutions in terms of system infrastructure, management and orchestration, and data-driven machine learning designs, thus realizing a new frontier for vehicular edge networks.

\section*{Acknowledgement}
The authors would like to acknowledge the support from the NCDOT under the award number TCE2020-03 and Cisco Systems, Inc.. The contents do not necessarily reflect the official views or policies of either the NCDOT or the Federal Highway Administration at the time of publication.
K.-C. Chen appreciates Cyber Florida to support part of this research.

\bibliographystyle{IEEEtran}
\bibliography{IEEEabrv,ds}

%\end{document}

\begin{IEEEbiography}%[{\includegraphics[width=1in,height=1.25in,clip,keepaspectratio]{scphoto.eps}}]
{Shih-Chun~Lin} [M'17] (slin23@ncsu.edu) received his Ph.D. degree in electrical and computer engineering from Georgia Institute of Technology in 2017. He is currently an assistant professor with the Department of Electrical and Computer Engineering, North Carolina State University, where he leads the Intelligent Wireless Networking Laboratory. His research interests include 5G and beyond, wireless software-defined architecture, vehicular edge computing and the artificial intelligence of things, machine learning and mathematical optimization, statistical scheduling and traffic engineering, and performance evaluation. He received the Distinguished TPC Member Award at IEEE INFOCOM 2020 and the Best Student Paper Award Runner-Up at IEEE SCC 2016.
\end{IEEEbiography}

\begin{IEEEbiography}%[{\includegraphics[width=1in,height=1.34in,clip,keepaspectratio]{n4.eps}}]
{Kwang-Cheng Chen} [F'07] (kwangcheng@usf.edu) is a professor of Electrical Engineering, University of South Florida. He has widely served in IEEE conference organization and journal editorship. He has contributed essential technology to IEEE 802, Bluetooth, LTE and LTE-A, and 5G-NR wireless standards. He has received a number of IEEE awards. His recent research interests include wireless networks, artificial intelligence and machine learning, IoT and CPS, social networks, and cybersecurity.
\end{IEEEbiography}

\begin{IEEEbiography}%[{\includegraphics[width=1in,height=1.34in,clip,keepaspectratio]{n4.eps}}]
{Ali Karimoddini} (akarimod@ncat.edu) received his Ph.D. degree from the National University of Singapore in 2012 and then joined the University of Notre Dame to conduct his postdoctoral studies. He is currently an associate professor with the Department of Electrical and Computer Engineering, North Carolina A\&T State University. He is also the Director of the NC-CAV Center of Excellence on Advanced Transportation, the Director of the ACCESS Laboratory, and the Deputy Director of the TECHLAV DoD Center of Excellence. His research interests include cyber-physical systems, control and robotics, resilient control systems, flight control systems, multiagent systems, and human–machine interactions. His research has been supported by different federal funding agencies and industrial partners. He is a member of AIAA, ISA, and AHS.
\end{IEEEbiography}

%\enlargethispage{-1.8in}

\end{document}